# An open dataset of scholars on Twitter

Philippe Mongeon
School of Information Management, Dalhousie University, Canada

Timothy D. Bowman
School of Information Sciences, Wayne State University, USA

Rodrigo Costas
Centre for Science and Technology Studies (CWTS), Leiden University, Leiden, The Netherlands; DSI-NRF Centre of Excellence in Scientometrics and Science, Technology and Innovation Policy (SciSTIP), Stellenbosch University, Stellenbosch, South Africa

## Abstract

The role played by research scholars in the dissemination of scientific knowledge on social media has always been a central topic in social media metrics (altmetrics) research. Different approaches have been implemented to identify and characterize active scholars on social media platforms like Twitter. Some limitations of past approaches were their complexity and, most importantly, their reliance on licensed scientometric and altmetric data. The emergence of new open data sources like OpenAlex or Crossref Event Data provides opportunities to identify scholars on social media using only open data. This paper presents a novel and simple approach to match authors from OpenAlex with Twitter users identified in Crossref Event Data. The matching procedure is described and validated with ORCID data. The new approach matches nearly 500,000 matched scholars with their Twitter accounts with a level of high precision and moderate recall. The dataset of matched scholars is described and made openly available to the scientific community to empower more advanced studies of the interactions of research scholars on Twitter.

## Introduction

Engagement with academic research on social media has been a central research topic in scientometrics, particularly in altmetrics and social media metrics. In the early days of social media metrics research, the focus was primarily on investigating the relationship between the number of mentions of research publications on social media platforms (particularly Twitter) and citations, with most of the studies finding weak relationships between social media metrics and citations (Costas et al., 2014; Sugimoto et al., 2017; Thelwall et al., 2013). However, recent theoretical proposals initiated a shift in the focus of altmetric research from analyzing mentions and correlations to more interactive perspectives. Thus, Haustein et al. (2016) proposed that social media metrics need not be restricted to the mentions of scholarly outputs on social media but could also include the mentions and activities of individual scholars. More recently, Costas et al. (2021) proposed the notion of "heterogeneous couplings" as a common framework to study the interactions between academic and non-academic actors as captured via online and social media platforms (see also Williams (2022)), in which the interactions of individual scholars on Twitter are another fundamental form of online interaction relating to how science is being communicated to society (Brainard, 2022).

In the quest to study scholars' activities on Twitter, one long-lasting challenge is the identification of social media accounts belonging to researchers. In 2020 we published a paper that introduced a method to match Web of Science authors with their Twitter accounts (Costas et al., 2020) and reported on the distribution of scholars on Twitter across countries, disciplines, academic age, and gender. One of the main features of that dataset was that it allowed, for the first time, to investigate the relationship between the research profiles and activities of scholars and their profiles and activities on Twitter on a large scale (Ferreira et al., 2021). Past datasets did not allow for this because they were either too small or because they provided information on whether or not a Twitter account *likely* belonged to a researcher without identifying the specific researcher to whom the account belonged.

Limitations of the dataset we produced with this initial work included that it used proprietary data from the Web of Science and Altmetric.com, which made it impossible to share the author-tweeter pairs openly, and it was complicated for others to use the dataset without access to these databases. The large number of steps involved in the previously reported process also possibly contributed to its lack of implementation by other researchers and its lack of transferability to other datasets.

New developments in Open Science scientometric and altmetric databases (namely the OpenAlex and Crossref Event Data databases) have changed this landscape, now allowing for the creation of matches of academic authors and their research publications with their Twitter profiles. This short paper aims to introduce a dataset of scholars' Twitter accounts identified with a naïve algorithm based entirely on available open data, presenting the process in detail with accompanying R and Python scripts so that the process can be easily replicated and/or improved upon by the research community. We hope this dataset will support further research on the interactions of scientific authors on social media and serve as a base for developing alternative and/or complementary approaches to match Twitter users and authors.

The paper is structured as follows. First, we provide an overview of the different data sources we used and the detailed process we used to match the Twitter accounts with OpenAlex authors. We then report precision and recall estimates for our matching approach, followed by an overview of the characteristics of the scholars found on Twitter.

## Data and methods

### Data sources

*OpenAlex*
The research publications data source for this study is the OpenAlex (Priem et al., 2022) data dump from 2022-05-20, which was downloaded and parsed into a relational database model hosted at the Maritime Institute for Science, Technology, and Society (MISTS) in Canada. In the OpenAlex database, authors are represented by a unique identifier (author_id) associated with their works (see the *works_authorships* table of the OpenAlex schema). Our ultimate objective is to assign a twitter_id to these author_id values. OpenAlex also includes a link between the author_id and ORCID. It is worth noting that a single individual can have multiple author_ids in OpenAlex, so that the same ORCID can be associated with multiple author_ids.

*Crossref Event Data*
We use a data dump of Crossref Event Data from January 2022 available at the Centre for Science and Technology Studies (CWTS), containing over 60 million Twitter events, which contain the tweet identifier and the DOI of the papers mentioned in that tweet. The dump includes 4.7 million unique DOIs tweeted at least once and recorded in Crossref Event Data (CED). We use the Twitter API to rehydrate the profile information of the Twitter users recorded in the CED dump. An important difference between CED and Altmetric is that CED focuses on identifying tweets to DOIs, while Altmetric also identifies Twitter mentions to preprints (e.g. from ArXiv) and other publication identifiers (e.g. PMIDs). Therefore, CED will typically identify fewer tweets to publications than Altmetric (Ortega, 2018).

Our use of an altmetric database like Crossref Event Data to identify researchers on Twitter stems from the expectation, as in Costas et al. (2020), that researchers are more likely to tweet research publications than non-researchers (Tsou et al., 2015) and therefore, to be recorded in this database. By considering only Twitter users that have mentioned scholarly work in their tweets as recorded in Crossref Event Data, we presumably increase precision at the expense of excluding all scholarly Twitter users that have never tweeted any research publications.

*ORCID*
The OpenAlex database includes the ORCID ids for approximately 2% of all the authors indexed in the database. Since some researchers include their Twitter handle in their public ORCID profiles, we leverage the information recorded in the ORCID Public Data File 2021 (Blackburn, Rob et al., 2021) to retrieve the Twitter account for those researchers who self-reported a Twitter profile in their ORCID profile. We used the ORCID data dump (2021) hosted by the Centre for Science and Technology Studies (CWTS) to obtain a set of 13,208 matching OpenAlex author ids and ORCID profiles. This dataset has been used as a golden set to evaluate the performance of our matching process. It should be noted that Twitter accounts listed in ORCID profiles are not necessarily valid. Even when these accounts are valid, it is important to note that the Twitter handle and user name may not include an actual name, which makes it impossible to match the accounts using the process described below as it relies on matching names. These factors may artificially penalize the recall, precision, and F-scores reported in the results section.

## Matching process
A central element in our current approach is the assumption that among the Twitter users tweeting a given publication, there will likely be one or more of the authors of that publication. Thus, this method is limited to identifying scholars on Twitter who tweeted (at least once) one of their publications (recorded on Crossref Event Data). This differs from the previous approach (Costas et al. 2020), which attempted to capture a broader range of relationships between Twitter users and authors to identify matches. This focus on "self-tweets" is likely to increase precision in matching authors and Twitter accounts but likely to decrease recall. However, we still expect to correctly identify a substantial number of authors on Twitter, as self-tweeting has been seen as an important form of researchers' engagement on Twitter (Ferreira et al., 2021). This focus on self-tweets also has the advantage of being less complex and computationally intensive, thus more easily implementable and replicable by others.

*Matching Twitter users with the authors of the tweeted papers*
To identify tweeter-author pairs, we take every tweeted paper and attempt to determine whether one of the authors' names matches the name of the Twitter user. An important feature of our approach to matching authors to Twitter users (similar to Costas et al. 2020) is that researchers must, to some degree, use a similar form of their name in their Twitter profile name. Our process does not aim to, and would not be able to, match the OpenAlex author id and the Twitter account of a researcher who uses a substantially different name in their Twitter and their authored works. For example, we will not match an author named Jane Smith using "squirl1" as their Twitter profile name.

However, although we require some similarity between author names and Twitter names, they can be recorded differently in Twitter and OpenAlex (and from one OpenAlex author record to another). Name variations can include the use of initials instead of the full first name, the inclusion/omission of middle names in full or initial form, and the inclusion of extra characters representing professional titles (e.g., Dr., Ph.D., MD). This requires normalizing the names from both data sources to maximize the likelihood that valid matches will be identified. Following the process used by Mongeon et al. (2017) to match dataset creators to Web of Science authors, we extract the last names(s), first name(s), and initial(s) of both Twitter users and OpenAlex authors and store them in distinct table columns. In those cases where a name contains more than two parts, we create an entry for all name combinations, assuming that all middle parts of the name can be part of the first or the last name. For instance, two entries would be created for the name John William Smith, one considering William as the second given name and one considering the token William as the first part of the last name. For each entry, we add a table column containing the initials (the first letter of each token that forms the first name), a table column containing the first initial only, and a table column containing the first token of the first name only.

The temporary tables used in the matching process include the unique ID of the individual (tweeter_id for Twitter and author_id for OpenAlex), the name, the deconstructed name variations, and the DOI tweeted in the event. Table 1 displays an example set of tweeter records.

Table 1. Example of a Twitter user record and extracted name components and variants.

| Tweeter_id | Handle | Profile name | First name | Last name | Initials | First initial | First token |
|---|---|---|---|---|---|---|---|
| 12345678 | jwsmith | John William Smith | John | William Smith | J | J | John |
| 12345678 | jwsmith | John William Smith | John William | Smith | JW | J | John |

We repeat the same process for OpenAlex authors and obtain a table like table 2.

Table 2. Example OpenAlex record and extracted name components and variants.

| Author_id | Display name | First name | Last name | Initials | First initial | First token |
|---|---|---|---|---|---|---|
| 12345678 | John William Smith | john | william smith | j | j | john |
| 12345678 | John William Smith | john william | smith | js | j | john |

We use different matching steps with different levels of expected precision and recall ranging from exact matches on the full Twitter profile name and Author display name (highest expected precision) to matches between the first initial and the last name (lowest expected precision). We perform the same set of steps using the profile name and the handle name. However, since the handle names are a single string without spaces, for those matching attempts, we concatenate the parts of the author's name in a single string as well. Table 3 presents a list of attempted matches, including example matches.

Table 3. Matching steps used to match Twitter profiles and OpenAlex author profiles.

| Step | Twitter data field | Examples |
|------|-------------------|----------|
| Full name exact match | Profile name | **john william smith = john william smith** |
| Full name substring* | Profile name | **john smith = john smith** Jr. |
| Last name + initials | Profile name | **jw smith = jw smith** |
| Last name + first token | Profile name | **john** w **smith = john smith** |
| Last name + first initial | Profile name | **jw smith = j smith** |
| Full name exact match | Handle | **john william smith = johnwilliamsmith** |
| Last name + initials | Handle | **jw smith = jwsmith** |
| Last name + first token | Handle | **john** w **smith = johnsmith** |
| Last name + first initial | Handle | **jw smith = jsmith** |

* Note that the full name substring step is only performed with the Twitter profile name since preliminary attempts to search for the full name as a substring of the Twitter handle performed extremely poorly (F-score < 0.1)

## Results

Our results are divided into two parts. First, we report on the performance of our matching algorithm using the recall, precision, and F-score based on our golden set of authors with an ORCID listed in OpenAlex and a Twitter handle in their ORCID account. In the second part, we describe our dataset by presenting the distribution of authors with Twitter accounts across fields and countries.

### Performance of the matching algorithm

We use the self-reported tweeter-author matches obtained from ORCID to evaluate the performance of the matching process at each step of the matching process. Precision is calculated by dividing the number of true positives by the total number of matches found for the tweeters in our golden set:

$$Precision = \frac{true\ positives}{true\ positives + false\ positives}$$

The recall is obtained by dividing the number of true positives by the total number of tweeter-author pairs in the golden set:

$$Recall = \frac{true\ positives}{true\ positives + false\ negatives}$$

The F-score is a measure of a model's accuracy on a dataset that is obtained with the following formula:

$$F = 2\ x\ \frac{precision\ x\ recall}{precision + recall}$$

All three indicators take a value between 0 and 1, where 1 is the best possible score.

Table 4 provides, for each of our matching criteria, the number of distinct matches obtained, the OpenAlex authors and Twitter accounts that form these matches, as well as the precision, recall, and F-scores obtained by testing each set of results against the golden set of author-tweeter pairs from ORCID.

Table 4. Results of the matching for each criterion.

| Matching step | | Distinct matches | | | Test | | |
|---|---|---|---|---|---|---|---|
| Criteria | Field | OpenAlex authors | Twitter accounts | Pairs | Recall | Precision | F-score |
| Last name + first token | Handle | 24,929 | 21,755 | 24,929 | 0.041 | 0.981 | 0.078 |
| Full name exact match | Handle | 19,147 | 16,795 | 19,147 | 0.033 | 0.979 | 0.063 |
| Last name + initials | Handle | 13,577 | 11,693 | 13,579 | 0.021 | 0.977 | 0.041 |
| Last name + first initial | Handle | 8,528 | 7,247 | 8,530 | 0.012 | 0.976 | 0.024 |
| Full name exact match | Profile name | 307,270 | 272,409 | 308,880 | 0.423 | 0.971 | 0.590 |
| Last name + first token | Profile name | 419,805 | 368,832 | 422,341 | 0.553 | 0.971 | 0.705 |
| Full name substring | Profile name | 317,723 | 281,499 | 319,984 | 0.442 | 0.968 | 0.607 |
| Last name + initials | Profile name | 343,649 | 299,646 | 346,529 | 0.458 | 0.967 | 0.621 |
| Last name + first initial | Profile name | 471,763 | 406,389 | 477,383 | 0.593 | 0.961 | 0.734 |
| **Combined** | **Combined** | **492,142** | **423,924** | **498,680** | **0.623** | **0.958** | **0.755** |

Because the same pairs can be obtained at different precision steps, we report in table 5 a different set of results where each step is performed hierarchically from most precise to least precise as per table 4, and where each step considers only new pairs that were not identified in the previous ones. This does not change the overall result of the matching but provides a more unambiguous indication of the contribution of each step to the recall and precision. For instance, we can see that the lowest precision rate (0.80) is obtained when matching the last name and first initial with the Tweeter's profile name. This is consistent with the test results presented in table 4. However, the difference in the precision between steps is more remarkable here because the count of true positives is not inflated by the valid matches from previous, more precise steps.

Table 5. New matches identified at each step of the hierarchical matching process.

| Matching step | | Distinct matches | | | Test | | |
|---|---|---|---|---|---|---|---|
| Criteria | Field | OpenAlex authors | Twitter accounts | Pairs | Recall | Precision | F-score |
| Last name + first token | Handle | 24,929 | 21,755 | 24,929 | 0.041 | 0.981 | 0.078 |
| Full name exact match* | Handle | 0 | 0 | 0 | 0 | 0 | 0 |
| Last name + initials | Handle | 13,325 | 11,513 | 13,327 | 0.020 | 0.976 | 0.040 |
| Last name + first initial | Handle | 1,976 | 1,758 | 1,976 | 0.003 | 0.966 | 0.006 |
| Full name exact match | Profile name | 283,862 | 51,936 | 285,373 | 0.380 | 0.971 | 0.546 |
| Last name + first token | Profile name | 96,587 | 87,508 | 97,088 | 0.101 | 0.964 | 0.182 |
| Full name substring | Profile name | 16,401 | 14,620 | 16,756 | 0.026 | 0.901 | 0.050 |
| Last name + initials | Profile name | 35,467 | 32,104 | 35,772 | 0.031 | 0.903 | 0.059 |
| Last name + first initial | Profile name | 22,989 | 20,815 | 23,459 | 0.017 | 0.806 | 0.033 |
| **Combined** | **Combined** | **492,142** | **423,924** | **98,680** | **0.623** | **0.958** | **0.755** |

\* All matches obtained during the full name matching with the tweeter handle were also found in the previous step, meaning that this step could, in principle, be skipped. However, we keep the step in our process for consistency and to account for the possibility that this step might yield new matches in future implementations of the process.

We can observe from Table 5 that the last three matching steps have a significantly lower precision rate than the other steps, especially the matches on the Last name + first initial between the author's name and the Tweeter's profile name. Researchers who might use our dataset or our process are thus advised to exercise caution with these less precise matching steps. While the results presented here do not include any manual data validation, we performed such a validation for the matches obtained with the three least precise steps. The dataset available on Zenodo (https://zenodo.org/record/7013518) includes a *validation* column alongside the tweeter_id and the author_id for the matches, which will allow users of the dataset to use the entire set or to filter out the matches that our team identified as likely to be false positives.

Table 6 presents yet another variation of the hierarchical matching process reported in table 5, with rows containing the cumulative matches so we can more clearly see how each step adds to the overall results.

Table 6. Cummulative number of matches at each step of the hierarchical matching process.

| Matching step | | Distinct matches | | | Test | | |
|---|---|---|---|---|---|---|---|
| Criteria | Field | OpenAlex authors | Twitter accounts | Pairs | Recall | Precision | F-score |
| Last name + first token | Handle | 24,929 | 21,755 | 24,929 | 0.041 | 0.981 | 0.078 |
| Full name exact match | Handle | 24,929 | 21,755 | 24,929 | 0.041 | 0.981 | 0.078 |
| Last name + initials | Handle | 38,248 | 33,201 | 38,256 | 0.061 | 0.979 | 0.115 |
| Last name + first initial | Handle | 40,223 | 34,841 | 40,232 | 0.064 | 0.979 | 0.120 |
| Full name exact match | Profile name | 323,935 | 286,528 | 325,605 | 0.445 | 0.972 | 0.611 |
| Last name + first token | Profile name | 420,167 | 369,257 | 422,693 | 0.548 | 0.970 | 0.700 |
| Last name + initials | Profile name | 436,170 | 382,820 | 439,449 | 0.574 | 0.967 | 0.720 |
| Full name substring | Profile name | 470,307 | 407,658 | 475,221 | 0.605 | 0.963 | 0.743 |
| Last name + first initial | Profile name | 492,142 | 423,924 | 498,680 | 0.623 | 0.958 | 0.755 |
| **Combined** | **Combined** | **492,142** | **423,924** | **498,680** | **0.623** | **0.958** | **0.755** |

## Overview of the dataset

In this section, we provide an overview of the composition of the dataset by looking at the discipline and countries of the researchers for which we were able to assign a Twitter account. There is no discipline classification in OpenAlex, but works are linked to Wikidata concepts (Priem et al., 2022), with a score ranging from 0 to 1 representing the strength of the association between the work and the concept. The concepts are hierarchical (levels 0 to 5), with level 0 essentially representing large disciplines (e.g., environmental science, economics, engineering, chemistry, medicine). For each unique author matched with a Twitter account, we retrieve their works, the level 0 concepts associated with these works, as well as the score. Figure 1 displays the sum of the scores for each concept. While these concepts are unlikely to provide a highly accurate classification of works, they are still helpful, we believe, in getting a sense of the breadth of disciplines represented in our dataset. Table 7 shows the number and percentage of authors assigned to each discipline based on the discipline with the highest score based on all their publications. Because authors are associated with each discipline to some degree (represented by the average score of concepts in their publications), we also display the average score for each discipline as an alternate representation of the distribution of disciplines in the dataset.

Table 7. Number of authors and the average score for by discipline (Concepts from OpenAlex).

| Discipline | Number of authors | Percentage of authors | Average score |
|---|---|---|---|
| Medicine | 138,968 | 28.3% | 0.517 |
| Biology | 75,246 | 15.3% | 0.388 |
| Psychology | 46,793 | 9.5% | 0.265 |
| Computer science | 39,675 | 8.1% | 0.201 |
| Political science | 36,531 | 7.4% | 0.229 |
| Chemistry | 30,876 | 6.3% | 0.242 |
| Materials science | 19,347 | 3.9% | 0.233 |
| Environmental science | 17,861 | 3.6% | 0.208 |
| Business | 17,485 | 3.6% | 0.158 |
| Sociology | 17,248 | 3.5% | 0.185 |
| Geography | 13,513 | 2.8% | 0.155 |
| Economics | 8,075 | 1.6% | 0.162 |
| Geology | 7,204 | 1.5% | 0.217 |
| Physics | 7,138 | 1.5% | 0.152 |
| Art | 6,963 | 1.4% | 0.133 |
| History | 3,913 | 0.8% | 0.134 |
| Philosophy | 2,415 | 0.5% | 0.097 |
| Mathematics | 1,637 | 0.3% | 0.068 |
| Engineering | 326 | 0.1% | 0.042 |
| **Total** | **491,214** | **100%** | |

For the countries, we use the *last_known_affiliation* field of the OpenAlex *authors* table and present the relative frequency of countries in Table 8.

Table 8. Distribution of authors' last known affiliation recorded in OpenAlex

| Country | Number of authors | Percentage of authors |
|---|---|---|
| United States | 142,059 | 32.1% |
| Great Britain | 72,430 | 16.4% |
| Australia | 24,457 | 5.5% |
| Canada | 22,516 | 5.1% |
| Spain | 19,308 | 4.4% |
| Germany | 18,338 | 4.1% |
| France | 10,794 | 2.4% |
| Netherlands | 10,605 | 2.4% |
| India | 9,967 | 2.3% |
| Italy | 9,016 | 2.0% |
| Brazil | 7,330 | 1.7% |
| Switzerland | 6,251 | 1.4% |
| Sweden | 5,684 | 1.3% |
| Ireland | 5,382 | 1.2% |
| Belgium | 5,375 | 1.2% |
| China | 5,250 | 1.2% |
| Finland | 4,834 | 1.1% |
| Denmark | 4,543 | 1.0% |
| Japan | 4,361 | 1.0% |
| Other countries | 54,093 | 12.2% |
| **Total** | **442,593** | **100%** |

# Discussion and conclusion

The work presented in this paper can be framed as a step forward in developing more advanced studies of the interactions between science and society, particularly by enabling the study of the role of scientists in disseminating scientific results on Twitter. In addition, using open data sources (OpenAlex and Crossref Event Data) allows researchers to continue to improve and adapt this method for further possibilities without worrying about contractual limitations or data unavailability.

Overall, the results of our matching process show a high level of precision and a moderate level of recall, which was expected given our consideration of only self-tweets in the matching process. This focus on self-tweets naturally makes a more precise matching strategy but at the expense of recall, since the method excludes from the matching all those scholars who never tweeted any of their publications (or none of their publications were included in the Crossref Event Data database).

The resulting matched database is more extensive than the one reported by Costas et al. (2020), which is most likely due to the broader coverage of OpenAlex compared to the Web of Science and improvements to the matching process. Other factors could include increasing Twitter use by researchers over time and/or increasing paper sharing practices on Twitter, or that more scholars are using their full names in their Twitter profiles.

It is important to emphasize the limitations of the matching approach (and resulting dataset) presented in this paper. First, the approach is limited to Tweets recorded in Crossref Event Data, which does not include Tweets before 2017, and publications and researchers recorded in OpenAlex. Furthermore, our matching algorithm requires that names be written in the Latin alphabet, which may exclude some Twitter users or authors who are not using the Latin alphabet. On top of this limitation in the coverage of our data sources, our reliance on self-tweets to increase the precision of our matches comes at the expense of some degree of recall. While we find that this strategy does provide high precision levels, our examination of the data signals that very common names can still generate some false positives. It may also create a gender bias in our dataset since a study by Peng et al. (2022) found men to be more likely to self-promote on Twitter than women. Finally, we also know from past research that there is a lower uptake of Twitter use in some regions or countries (Zahedi, 2017).

The characterization of the social media users and audiences that are engaging with scientific publications is an important element in the development of more advanced studies on the interactions between science and society. Thus, the further and better curation of data around the interactions between social media users and academic objects is a fundamental step that needs to be considered in future altmetric research. Further developments of this work include expanding the matching to include additional matching criteria and developing approaches (e.g., citation or co-author networks) to form tweeter-author pairs without relying only on self-tweets. Future developments could also tackle the challenge of matching scholars with Twitter users that did not tweet any publication. Nevertheless, we believe that the relatively simple matching process outlined in this paper and the open dataset of nearly half a million pairs of OpenAlex author IDs and Tweeter IDs generated and made available to the community are valuable contributions to the field of quantitative science studies, and more specifically to the study of the activities of scholars on Twitter and the interaction between social media and science.

## Acknowledgements

The authors would like to thank Poppy Riddle, Kydra Mayhew, and Maddie Hare, who helped with data cleaning. Rodrigo Costas is partially funded by the South African DSI-NRF Centre of Excellence in Scientometrics and Science, Technology and Innovation Policy (SciSTIP).

## Supplementary materials

The dataset produced with the process reported in this paper is available at
https://zenodo.org/record/7013518